# The Turing Trap:

## The Promise & Peril of Human-Like Artificial Intelligence

Erik Brynjolfsson

Stanford Digital Economy Lab

http://brynjolfsson.com

*In 1950, Alan Turing proposed an "imitation game" as the ultimate test of whether a machine was intelligent: could a machine imitate a human so well that it's answers to questions indistinguishable from a human's.[1] Ever since, creating intelligence that matches human intelligence has implicitly or explicitly been the goal of thousands of researchers, engineers and entrepreneurs. The benefits of human-like artificial intelligence (HLAI) include soaring productivity, increased leisure, and perhaps most profoundly, a better understanding of our own minds.*

*But not all types of AI are human-like – in fact, many of the most powerful systems are very different from humans – and an excessive focus on developing and deploying HLAI can lead us into a trap. As machines become better substitutes for human labor, workers lose economic and political bargaining power and become increasingly dependent on those who control the technology. In contrast, when AI is focused on augmenting humans rather than mimicking them, then humans retain the power to insist on a share of the value created. What's more, augmentation creates new capabilities and new products and services, ultimately generating far more value than merely human-like AI. While both types of AI can be enormously beneficial, there are currently excess incentives for automation rather than augmentation among technologists, business executives, and policy-makers.*

Alan Turing was far from the first to imagine human-like machines. According to legend, 3,500 years ago, Dædalus constructed humanoid statues that were so lifelike that they moved and spoke by themselves.[2] Nearly every culture has its own stories of human-like machines, from Yanshi's leather man described in the ancient Chinese *Liezi* text to the bronze Talus of the Argonautica and the towering clay *Mokkerkalfe* of Norse mythology. The word robot first appeared in Karel Čapek's influential play *Rossum's Universal Robots* and derives from the Czech word *robota*, meaning servitude or work. In fact, in the first drafts of his play, Čapek named them *labori* until his brother Josef suggested substituting the word robot.[3]

Of course, it is one thing to tell tales about humanoid machines. It is something else to create robots that do real work. For all our ancestors' inspiring stories, we are the first generation to build and deploy real robots in large numbers.[4] Dozens of companies are working on robots as human-like, if not more so, as those described in the ancient texts. One might say that technology has advanced sufficiently to become indistinguishable from mythology.[5]

The breakthroughs in robotics depend not merely on more dexterous mechanical hands and legs, and more perceptive synthetic eyes and ears, but also on increasingly human-like artificial intelligence. Powerful AI systems are crossing key thresholds: matching humans in a growing number of fundamental tasks such as image recognition and speech recognition, with applications from autonomous vehicles and medical diagnosis to inventory management and product recommendations.[6] AI is appearing in more and more products and processes.[7]

These breakthroughs are both fascinating and exhilarating. They also have profound economic implications. Just as earlier general-purpose technologies like the steam engine and electricity catalyzed a restructuring of the economy, our own economy is increasingly transformed by AI. A good case can be made that AI is the most general of all general-purpose technologies: after all, if we can solve the puzzle of intelligence, it would help solve many of the other problems in the world,. And we are making remarkable progress. In the coming decade, machine intelligence will become increasingly powerful and pervasive. We can expect record wealth creation as a result.

Replicating human capabilities is valuable not only because of its practical potential for reducing the need for human labor, but also because it can help us build more robust and flexible forms of intelligence. Whereas domain-specific technologies can often make rapid progress on narrow tasks, they founder when unexpected problems or unusual circumstances arise. That is where human-like intelligence excels. In addition, HLAI could help us understand more about ourselves. We appreciate and comprehend the human mind better when we work to create an artificial one.

Let's look more closely at how HLAI could lead to a realignment of economic and political power.

The distributive effects of AI depend on whether it is primarily used to augment human labor or automate and replace it. When AI augments human capabilities, enabling people to do things they never could before, then humans and machines are complements. Complementarity implies that people remain indispensable for value creation and retain bargaining power in labor markets and in political decision-making. In contrast, when AI replicates and automates existing human capabilities, machines become better substitutes for human labor and workers lose economic and political bargaining power. Entrepreneurs and executives who have access to machines with capabilities that replicate those of human for a given task can and often will replace humans in those tasks.

A fully automated economy could, in principle, be structured to redistribute the benefits from production widely, even to those who are no longer strictly necessary for value creation. However, the beneficiaries would be in a weak bargaining position to prevent a change in the distribution that left them with little or nothing. They would depend precariously on the decisions of those in control of the technology. This opens the door to increased concentration of wealth and power.

This highlights the promise and the peril of achieving HLAI: building machines designed to pass the Turing Test and other, more sophisticated metrics of human-like intelligence.[8] On the one hand, it is a path to unprecedented wealth, increased leisure, robust intelligence, and even a better understanding of ourselves. On the other hand, if HLAI leads machines to automate rather than augment human labor, it creates the risk of concentrating wealth and power. And with that concentration comes the peril of being trapped in an equilibrium where those without power have no way to improve their outcomes, a situation I call the *Turing Trap*.

The grand challenge of the coming era will be to reap the unprecedented benefits of AI, including its human-like manifestations, while avoiding the Turing Trap. Succeeding in this task requires an understanding of how technological progress affects productivity and inequality, why the Turing Trap is so tempting to different groups, and a vision of how we can do better.

\*\*\*

AI pioneer Nils Nilsson noted that "achieving real human-level AI would necessarily imply that most of the tasks that humans perform for pay could be automated."[9] In the same article, he called for a focused effort to create such machines, writing that "achieving human-level AI or 'strong AI' remains the ultimate goal for some researchers" and he contrasted this with "weak AI," which seeks to "build machines that help humans."[10] Not surprisingly, given these monikers, work toward "strong AI" attracted many of the best and brightest minds to the quest of–implicitly or explicitly–fully automating human labor, rather than assisting or augmenting it.

For the purposes of this essay, rather than strong versus weak AI, let us use the terms *automation* versus *augmentation*. In addition, I will use HLAI to mean human-*like* artificial intelligence not human-*level* AI because the latter mistakenly implies that intelligence falls on a single dimension, and perhaps even that humans are at the apex of that metric. In reality, intelligence is multidimensional: a 1970s pocket calculator surpasses the most intelligent human in some ways (such as multiplication), as does a chimpanzee (short-term memory). At the same time, machines and animals are inferior to human intelligence on myriad other dimensions. The term "artificial general intelligence" (AGI) is often used as a synonym for HLAI. However, taken literally, it is the union of all types of intelligences, able to solve types of problems that are solvable by any existing human, animal, or machine. That suggests that AGI is not human-like.

The good news is that both automation and augmentation can boost labor productivity: that is, the ratio of value-added output to labor-hours worked. As productivity increases, so do average incomes and living standards, as do our capabilities for addressing challenges from climate change and poverty to health care and longevity.[11] Mathematically, if the human labor used for a given output declines toward zero, then labor productivity would grow to infinity.[12]

The bad news is that no economic law ensures everyone will share this growing pie. Although pioneering models of economic growth[13] [14] assumed that technological change was neutral, in practice technological change can disproportionately help or hurt some groups, even if it is beneficial on average.[15]

In particular, the way the benefits of technology are distributed depends to a great extent on how the technology is deployed and the economic rules and norms that govern the equilibrium allocation of goods, services, and incomes. When technologies automate human labor, they tend to reduce the marginal value of workers' contributions, and more of the gains go to the owners, entrepreneurs, inventors, and architects of the new systems. In contrast, when technologies augment human capabilities, more of the gains go to human workers.[16]

A common fallacy is to assume that all or most productivity-enhancing innovations belong in the first category: automation. However, the second category, augmentation, has been far more important throughout most of the past two centuries. One metric of this is the economic value of an hour of human labor. Its market price as measured by median wages has grown more than ten-fold since 1820.[17] An entrepreneur is willing to pay much more for a worker whose capabilities are amplified by a bulldozer than one who can only work with a shovel, let alone with bare hands.

In many cases, not only wages but also employment grow with the introduction of new technologies. With the invention of jet engines, pilot productivity (in passenger-miles per pilot-hour) grew immensely. Rather than reducing the number of employed pilots, the technology spurred demand for air travel so much that the number of pilots grew. Although this pattern is comforting, past performance does not guarantee future results. Modern technologies–and, more important, the ones under development–are different from those that were important in the past.[18]

In recent years, we have seen growing evidence that not only is the labor share of the economy declining, but even among workers, some groups are beginning to fall even farther behind.[19] Over the past forty years, the numbers of millionaires and billionaires grew but the average real wages for Americans with only a high school education fell.[20] While many phenomena contributed to this, including new patterns of global trade, changes in technology deployment are the single biggest explanation.

If capital in the form of AI can perform more tasks, those with unique assets, talents, or skills that are not easily replaced with technology stand to benefit disproportionately.[21] The result has been greater wealth concentration.[22]

Ultimately, a focus on more human-like AI can make technology a better substitute for the many non-superstar workers, driving down their market wages, even as it amplifies the market power of a few.[23] This has created a growing fear that AI and related advances will lead to a burgeoning class of unemployable or "zero marginal product" people.[24]

\*\*\*

An unfettered market is likely to create socially excessive incentives for innovations that automate human labor and produce weak incentives for technology that augments humans. The first fundamental welfare theorem of economics states that under a particular set of conditions, market prices lead to a *pareto optimal* outcome: that is, one where no one can be made better off without making someone else worse off. But the theorem does not hold when there are innovations that change the production possibilities set or externalities that affect people who are not part of the market.

Both innovations and externalities are of central importance to the economic effects of AI, since AI is not only an innovation itself, but also one that triggers cascades of complementary innovations, from new products to new production systems.[25] Furthermore, the effects of AI, particularly on work, are rife with externalities. When a worker loses opportunities to earn labor income, the costs go beyond the newly unemployed to affect many others in their community and in the broader society. With fading opportunities often come the dark horses of alcoholism, crime, and opioid abuse. Recently, the United States has experienced the first decline in life expectancies in its recorded history, a result of increasing deaths from suicide, drug overdose, and alcoholism, what economists Anne Case and Angus Deaton call "deaths of despair."[26]

This spiral of marginalization can grow because concentration of economic power often begets concentration of political power. In the words attributed to Louis Brandeis: "We may have democracy, or we may have wealth concentrated in the hands of a few, but we can't have both." In contrast, when humans are indispensable to value creation, economic power will tend to be more decentralized. Historically, most economically valuable knowledge–what economist Simon Kuznets called "useful knowledge"–resided within human brains.[27] But no human brain can contain even a small fraction of the useful knowledge needed to run even a medium-sized business, let alone a whole industry or economy, so knowledge had to be distributed and decentralized.[28] The decentralization of useful knowledge, in turn, decentralizes economic and political power.

Unlike nonhuman assets such as property and machinery, much of a person's knowledge is inalienable, both in the practical sense that no one person can know everything that another person knows and in the legal sense that its ownership cannot be legally transferred.[29] In contrast, when knowledge becomes codified and digitized, it can be owned, transferred, and concentrated very easily. Thus, when knowledge shifts from humans to machines, it opens the possibility of concentration of power. When historians look back on the first two decades of the twenty-first century, they will note the striking growth in the digitization and codification of information and knowledge.[30] In parallel, machine learning models are becoming larger, with hundreds of billions of parameters, using more data and getting more accurate results.[31]

More formally, incomplete contracts theory shows how ownership of key assets provides bargaining power in relationships between economic agents (such as employers and employees, or business owners and subcontractors).[32] To the extent that a person controls an indispensable asset (like useful knowledge) needed to create and deliver a company's products and services, that person can command not only higher income but also a voice in decision-making. When useful knowledge is inalienably locked in human brains, so too is the power it confers. But when it is made alienable, it enables greater concentration of decision-making and power.[33]

***

The risks of the Turing trap are amplified because three groups of people–technologists, businesspeople, and policy-makers–each find it alluring. Technologists have sought to replicate human intelligence for decades to address the recurring challenge of what computers could not do. The invention of computers and the birth of the term "electronic brain" were the latest fuel for the ongoing battle between technologists and humanist philosophers.[34] The philosophers posited a long list of ordinary and lofty human capacities that computers would never be able to do. No machine could play checkers, master chess, read printed words, recognize speech, translate between human languages, distinguish images, climb stairs, win at Jeopardy or Go, write poems, and so forth.

For professors, it is tempting to assign such projects to their graduate students. Devising challenges that are new, useful, and achievable can be as difficult as solving them. Rather than specify a task that neither humans nor machines have ever done before, why not ask the research team to design a machine that replicates an existing human capability? Unlike more ambitious goals, replication has an existence proof that such tasks are, in principle, feasible and useful. While the appeal of human-like systems is clear, the paradoxical reality is that HLAI can be more difficult and less valuable than systems that achieve superhuman performance.

In 1988, robotics developer Hans Moravec noted[35] that "it is comparatively easy to make computers exhibit adult level performance on intelligence tests or playing checkers, and difficult or impossible to give them the skills of a one-year-old when it comes to perception and mobility." But I would argue that in many domains, Moravec was not nearly ambitious enough. It is often comparatively easier for a machine to achieve *superhuman* performance in new domains than to match ordinary humans in the tasks they do regularly.

Humans have evolved over millions of years to be able to comfort a baby, navigate a cluttered forest, or pluck the ripest blueberry from a bush, tasks that are difficult if not impossible for current machines. But machines excel when it comes to seeing X-rays, etching millions of transistors on a fragment of silicon, or scanning billions of webpages to find the most relevant one,. Imagine how feeble and limited our technology would be if past engineers set their sights on merely replicating human-levels of perception, actuation, and cognition. Augmenting humans with technology opens an endless frontier of new abilities and opportunities. The set of tasks that humans and machines can do together is undoubtedly much larger than those humans can do alone (Figure 1). Machines can perceive things that are imperceptible to humans, they can act on objects in ways that no human can, and they can comprehend things that are incomprehensible to the human brain. As Demis Hassabis, CEO of Deepmind, put it, the AI system "doesn't play like a human, and it doesn't play like a program. It plays in a third, almost alien, way . . . it's like chess from another dimension."[36] Computer scientist Jonathan Schaeffer explains the source of its superiority: "I'm absolutely convinced it's because it hasn't learned from humans."[37] More fundamentally, inventing tools that augment the process of invention itself promises to expand not only our collective abilities, but to accelerate the rate of expansion of those abilities.

Figure 1

[Labor Automation and Augmentation]

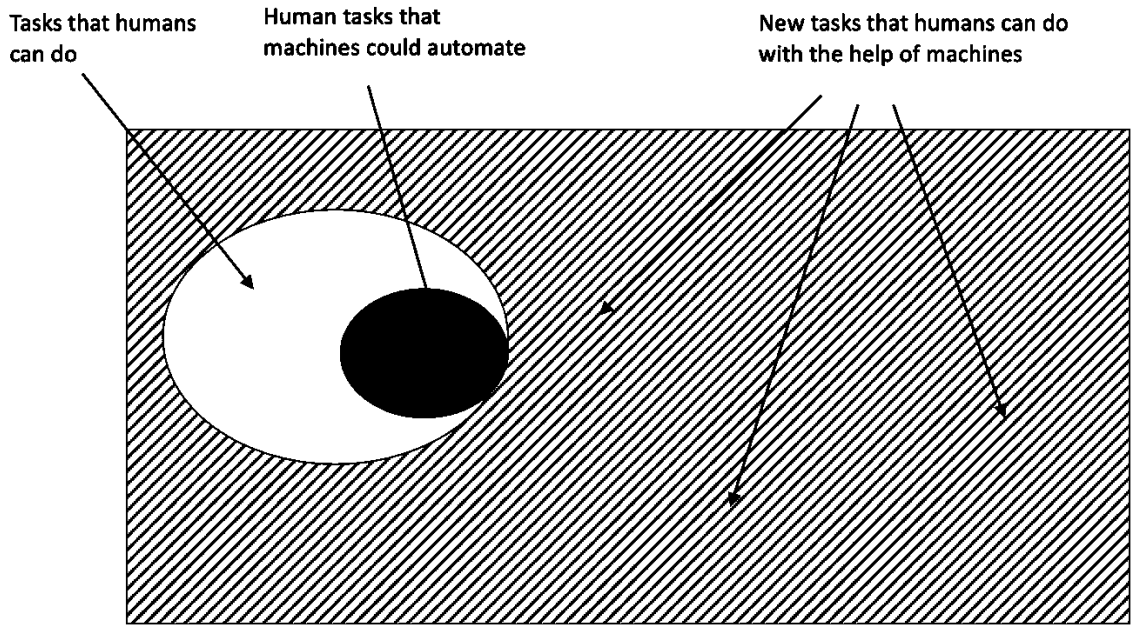

Figure 1: There is far more opportunity in augmenting humans to do new tasks, rather than automating what they can already do

What about businesspeople? They often find that substituting machinery for human labor is the low-hanging fruit of innovation. The simplest approach is to implement plug-and-play automation: swap in a piece of machinery for each task a human is currently doing. That mindset reduces the need for more radical changes to business processes.[38] Task-level automation reduces the need to understand subtle interdependencies and creates easy A-B tests, by focusing on a known task with easily measurable performance improvement.

Similarly, because labor costs are the biggest line item in almost every company's budget, automating jobs is a popular strategy for managers. Cutting costs–which can be an internally coordinated effort–is often easier than expanding markets. Moreover, many investors prefer "scalable" business models, which is often a synonym for a business that can grow without hiring and the complexities that entails.

But here again, when businesspeople focus on automation, they often set out to achieve a task that is both less ambitious and more difficult than it need be. To understand the limits of substitution-oriented automation, consider a thought experiment. What if our old friend Dædalus had at his disposal an extremely talented team of engineers 3,500 years ago and had, somehow, built human-like machines that fully automated every work-related task that his fellow Greeks were doing.

- Herding sheep? Automated.
- Making clay pottery? Automated.
- Weaving tunics? Automated.
- Repairing horse-drawn carts? Automated.
- Bloodletting victims of disease? Automated.

The good news is that labor productivity would soar, freeing the ancient Greeks for a life of leisure. The bad news is that their living standards and health outcomes would come nowhere near matching ours. After all, there is only so much value one can get from clay pots and horse-drawn carts, even with unlimited quantities and zero prices.

In contrast, most of the value that our economy has created since ancient times comes from new goods and services that not even the kings of ancient empires had, not from cheaper versions of existing goods.[39] In turn, myriad new tasks are required: fully 60 percent of people are now employed in occupations that did not exist in 1940.[40] In short, automating labor ultimately unlocks less value than augmenting it to create something new.

At the same time, automating a whole job is often brutally difficult. Most jobs involve many tasks that are extremely challenging to automate, even with the most clever technologies. For example, AI may be able to read mammograms better than a human radiologist, but it cannot do the other twenty-six tasks associated with the job, according to O-NET, such as comforting a concerned patient or coordinating on a care plan with other doctors.[41] My work with Tom Mitchell and Daniel Rock on the suitability for machine learning found many occupations in which machines could contribute some tasks, but zero occupations out of 950 in which machine learning could do 100 percent of the necessary tasks.[42]

The same principle applies to the more complex production systems that involve multiple people working together.[43] To be successful, firms typically need to adopt a new technology as part of a system of mutually reinforcing organizational changes.[44]

Consider another thought experiment: Imagine if Jeff Bezos had "automated" existing bookstores by simply replacing all the human cashiers with robot cashiers. That might have cut costs a bit, but the total impact would have been muted. Instead, Amazon reinvented the concept of a bookstore by combining humans and machines in a novel way. As a result, they offer vastly greater product selection, ratings, reviews, and advice, and enable 24/7 retail access from the comfort of customers' homes. The power of the technology was not in automating the work of humans in the existing retail bookstore concept but in reinventing and augmenting how customers find, assess, purchase, and receive books and, in turn, other retail goods.

Third, policy-makers have also often tilted the playing field toward automating human labor rather than augmenting it. For instance, the U.S. tax code currently encourages capital investment over investment in labor through effective tax rates that are much higher on labor than on plant and equipment.[45]

Consider a third thought experiment: two potential ventures each use AI to create one billion dollars of profits. If one of them achieves this by augmenting and employing a thousand workers, the firm will owe corporate and payroll taxes, while the employees will pay income taxes, payroll taxes, and other taxes. If the second business has no employees, the government may collect the same corporate taxes, but no payroll taxes and no taxes paid by workers. As a result, the second business model pays far less in total taxes.

This disparity is amplified because the tax code treats labor income more harshly than capital income. In 1986, top tax rates on capital income and labor income were equalized in the United States, but since then, successive changes have created a large disparity, with the 2021 top marginal federal tax rates on labor income of 37 percent, while long capital gains have a variety of favorable rules, including a lower statutory tax rate of 20 percent, the deferral of taxes until capital gains are realized, and the "step-up basis" rule that resets capital gains to zero, wiping out the associated taxes, when assets are inherited.

The first rule of tax policy is simple: you tend to get less of whatever you tax. Thus, a tax code that treats income that uses labor less favorably than income derived from capital will favor automation over augmentation. Undoing this imbalance would lead to more balanced incentives.

In fact, given the positive externalities of more widely shared prosperity, a case could be made for treating wage income *more* favorably than capital income, for instance by expanding the earned income tax credit.[46]

Government policy in other areas could also do more to steer the economy clear of the Turing Trap. The growing use of AI, even if only for complementing workers, and the further reinvention of organizations around this new general-purpose technology implies a great need for worker training or retraining. In fact, for each dollar spent on machine learning technology, companies may need to spend nine dollars on intangible human capital.[47] However, training suffers from a serious externality issue: companies that incur the costs to train or retrain workers may reap only a fraction of the benefits of those investments, with the rest potentially going to other companies, including competitors, as these workers are free to bring their skills to their new employers. At the same time, workers are often cash- and credit-constrained, limiting their ability to invest in their own skills development.[48] This implies that governments policy should directly provide this training or provide incentives for corporate training that offset the externalities created by labor mobility.[49]

In sum, the risks of the Turing Trap are increased not by just one group in our society, but by the misaligned incentives of technologists, businesspeople, and policy-makers.

***

The future is not preordained. We control the extent to which AI either expands human opportunity through augmentation or replaces humans through automation. We can work on challenges that are easy for machines and hard for humans, rather than hard for machines and easy for humans. The first option offers the opportunity of growing and sharing the economic pie by augmenting the workforce with tools and platforms. The second option risks dividing the economic pie among an ever-smaller number of people by creating automation that displaces ever-more types of workers.

While both approaches can and do contribute to progress, too many technologists, businesspeople, and policy-makers have been putting a finger on the scales in favor of replacement. Moreover, the tendency of a greater concentration of technological and economic power to beget a greater concentration of political power risks trapping a powerless majority into an unhappy equilibrium: the Turing Trap.

The backlash against free trade offers a cautionary tale. Economists have long argued that free trade and globalization tend to grow the economic pie through the power of comparative advantage and specialization. They have also acknowledged that market forces alone do not ensure that every person in every country will come out ahead. So they proposed a grand bargain: maximize free trade to maximize wealth creation and then distribute the benefits broadly to compensate any injured occupations, industries, and regions. It hasn't worked as they had hoped. As the economic winners gained power, they reneged on the second part of the bargain, leaving many workers worse off than before.[50] The result helped fuel a populist backlash that led to import tariffs and other barriers to free trade. Economists wept.

Some of the same dynamics are already underway with AI. More and more Americans, and indeed workers around the world, believe that while the technology may be creating a new billionaire class, it is not working for them. The more technology is used to replace rather than augment labor, the worse the disparity may become, and the greater the resentments that feed destructive political instincts and actions. More fundamentally, the moral imperative of treating people as ends, and not merely as means, calls for everyone to share in the gains of automation.

The solution is not to slow down technology, but rather to eliminate or reverse the excess incentives for automation over augmentation. In concert, we must build political and economic institutions that are robust in the face of the growing power of AI. We can reverse the growing tech backlash by creating the kind of prosperous society that inspires discovery, boosts living standards, and offers political inclusion for everyone. By redirecting our efforts, we can avoid the Turing Trap and create prosperity for the many, not just the few.

Author's note: The core ideas in this essay were inspired by a series of conversations with James Manyika and Andrew McAfee. I am grateful for valuable comments and suggestions on this work from Matt Beane, Seth Benzell, Katya Klinova, Alena Kykalova, Gary Marcus, Andrea Meyer, and Dana Meyer, but they should not be held responsible for any errors or opinions in the essay.

**Erik Brynjolfsson** is the Jerry Yang and Akiko Yamazaki Professor and Senior Fellow at the Institute for Human-Centered AI and Director of the Digital Economy Lab at Stanford University. He is also the Ralph Landau Senior Fellow at the Institute for Economic Policy Research and Professor by Courtesy at the Graduate School of Business and Department of Economics at Stanford University, and a Research Associate at the National Bureau of Economic Research. He is the author or co-author of seven books including (with Andrew McAfee): *Machine, Platform, Crowd: Harnessing Our Digital Future* (2017), *The Second Machine Age: Work, Progress, and Prosperity in a Time of Brilliant Technologies* (2014), and *Race against the Machine: How the Digital Revolution Is Accelerating Innovation, Driving Productivity, and Irreversibly Transforming Employment and the Economy* (2011) and (with Adam Saunders): *Wired for Innovation: How Information Technology Is Reshaping the Economy (2009).*

[1] Alan Turing (October 1950), "Computing Machinery and Intelligence", Mind, LIX (236): 433–460, doi:10.1093/mind/LIX.236.433. An earlier articulation of this test comes from Descartes in *The Discourse*, in which he wrote,

> If there were machines which bore a resemblance to our bodies and imitated our actions as closely as possible for all practical purposes, we should still have two very certain means of recognizing that they were not real men. The first is that they could never use words, or put together signs, as we do in order to declare our thoughts to others. . . . Secondly, even though some machines might do some things as well as we do them, or perhaps even better, they would inevitably fail in others, which would reveal that they are acting not from understanding.

[2] Carolyn Price, "Plato, Opinions and the Statues of Daedalus," OpenLearn, updated June 19, 2019, https://www.open.edu/openlearn/history-the-arts/philosophy/plato-opinions-and-the-statues-daedalus; and Andrew Stewart, "The Archaic Period," Perseus Digital Library, http://www.perseus.tufts.edu/hopper/text?doc=Perseus:text:1999.04.0008:part=2:chapter=1&highlight=daedalus.

[3] "The Origin of the Word 'Robot,'" Science Friday, April 22, 2011, https://www.sciencefriday.com/segments/the-origin-of-the-word-robot/.

[4] Millions of people are now working alongside robots. For a recent survey on the diffusion of robots, AI, and other advanced technologies in the United States, see Nikolas Zolas, Zachary Kroff, Erik Brynjolfsson, et al., "Advanced Technologies Adoption and Use by U.S. Firms: Evidence from the Annual Business Survey," NBER Working Paper No. 28290 (Cambridge, Mass.: National Bureau of Economic Research, 2020).

[5] Apologies to Arthur C. Clarke.

[6] See, for example, Daniel Zhang, Saurabh Mishra, Erik Brynjolfsson, et al., "The AI Index 2021 Annual Report," arXiv preprint arXiv:2103.06312 (Ithaca, N.Y.: Cornell University, 2021), esp.

chap. 2. In regard to image recognition, see, for instance, the success of image recognition systems in Olga Russakovsky, Jia Deng, Hao Su, et al., "Imagenet Large Scale Visual Recognition Challenge," *International Journal of Computer Vision* 115 (3) (2015): 211–252.

[7] Erik Brynjolfsson and Andrew McAfee, "The Business of Artificial Intelligence," *Harvard Business Review* (2017): 3–11.

[8] See for example, Hubert Dreyfus, *What Computers Can't Do* (Cambridge, Mass.: MIT Press, 1972), Nils J. Nilsson, "Human-Level Artificial Intelligence? Be Serious!" *AI Magazine* 26 (4) (2005): 68; and Gary Marcus, Francesca Rossi, and Manuela Veloso, "Beyond the Turing Test," *AI Magazine* 37 (1) (2016): 3–4.

[9] Nilsson, "Human-Level Artificial Intelligence?" 68.

[10] John Searle was the first to use the terms strong AI and weak AI, writing that with weak AI, "the principal value of the computer . . . is that it gives us a very powerful tool," while strong AI "really is a mind." Ed Feigenbaum has argued that creating such intelligence is the "manifest destiny" of computer science. (John R. Searle. 1980. Minds, Brains, and Programs. Behavioral and Brain Sciences 3(3): 417–57.

[11] If working hours fall fast enough, it is theoretically possible, though empirically unlikely, that living standards could fall even as productivity rises.

[12] However, as discussed below, this does not necessarily mean living standards would rise without bound.

[13] See, for example, Robert M. Solow, "A Contribution to the Theory of Economic Growth," *The Quarterly Journal of Economics* 70 (1) (1956): 65–94.

[15] See for example Daron Acemoglu, "Directed Technical Change," *Review of Economic Studies* 69 (4) (2002): 781–809.

[16] See, for instance, Erik Brynjolfsson and Andrew McAfee, *Race Against the Machine: How the Digital Revolution Is Accelerating Innovation, Driving Productivity, and Irreversibly Transforming Employment and the Econom*y (Lexington, Mass.: Digital Frontier Press, 2011); and Daron Acemoglu and Pascual Restrepo, "The Race Between Machine and Man: Implications of Technology for Growth, Factor Shares, and Employment," *American Economic Review* 108 (6) (2018): 1488–1542.

[17] For instance, the real wage of a building laborer in Great Britain is estimated to have grown from sixteen times the amount needed for subsistence in 1820 to 167 times that level by the year 2000, according to Jan Luiten Van Zanden, Joerg Baten, Marco Mira d'Ercole, et al., eds., *How Was Life? Global Well-Being since 1820* (Paris: OECD Publishing, 2014).

[18] For instance, a majority of aircraft on US Navy aircraft carriers are likely to be unmmaned. See Oriana Pawlyk, "Future Navy Carriers Could Have More Drones Than Manned Aircraft, Admiral Says", *Military.com*, March 30, 2021.
"
[19] Loukas Karabarbounis and Brent Neiman, "The Global Decline of the Labor Share," *The Quarterly Journal of Economics* 129 (1) (2014): 61–103; and David Autor, "Work of the Past, Work of the Future," NBER Working Paper No. 25588 (Cambridge, Mass.: National Bureau of Economic Research, 2019). For a broader survey, see Morgan R. Frank, David Autor, James E. Bessen, et al., "Toward Understanding the Impact of Artificial Intelligence on Labor," *Proceedings of the National Academy of Sciences* 116 (14) (2019): 6531–6539.

[20] Daron Acemoglu and David Autor, "Skills, Tasks and Technologies: Implications for Employment and Earnings," *Handbook of Labor Economics* 4 (2011): 1043–1171.

[21] Seth G. Benzell and Erik Brynjolfsson, "Digital Abundance and Scarce Architects: Implications for Wages, Interest Rates, and Growth," NBER Working Paper No. 25585 (Cambridge, Mass.: National Bureau of Economic Research, 2021).

[22] Prasanna Tambe, Lorin Hitt, Daniel Rock, and Erik Brynjolfsson, "Digital Capital and Superstar Firms," Hutchins Center Working Paper #73 (Washington, D.C.: Hutchins Center at Brookings, 2021), https://www.brookings.edu/research/digital-capital- and-superstar-firms.

[23] There is some evidence that capital is already becoming an increasingly good substitute for labor. See, for instance, the discussion in Michael Knoblach and Fabian Stöckl, "What Determines the Elasticity of Substitution between Capital and Labor? A Literature Review," *Journal of Economic Surveys* 34 (4) (2020): 852.

[24] See, for example, Tyler Cowen, *Average Is Over: Powering America beyond the Age of the Great Stagnation* (New York: Penguin, 2013). Or more provocatively, Yuval Noah Harari, "The Rise of the Useless Class," Ted Talk, February 24, 2017, https://ideas.ted.com/the-rise-of-the-useless-class/.

[25] Erik Brynjolfsson and Andrew McAfee, "Artificial Intelligence, for Real," *Harvard Business Review*, August 7, 2017.

[26] Robert D. Putnam, *Our Kids: The American Dream in Crisis* (New York: Simon and Schuster, 2016) describes the negative effects of joblessness, while Anne Case and Angus Deaton, *Deaths of Despair and the Future of Capitalism* (Princeton, N.J.: Princeton University Press, 2021) documents the sharp decline in life expectancy among many of the same people.

[27] Simon Smith Kuznets, *Economic Growth and Structure: Selected Essays* (New York: W. W. Norton & Co., 1965).

[28] Friedrich August Hayek, "The Use of Knowledge in Society," *The American Economic Review* 35 (4) (1945): 519–530.

[29] Erik Brynjolfsson, "Information Assets, Technology and Organization," *Management Science* 40 (12) (1994): 1645–1662, https:// doi.org/10.1287/mnsc.40.12.1645.

[30] For instance, in the year 2000, an estimated 85 billion (mostly analog) photos were taken, but by 2020, that had grown nearly twenty-fold to 1.4 trillion (almost all digital) photos.

[31] Andrew Ng, "What Data Scientists Should Know about Deep Learning," speech presented at Extract Data Conference, November 24, 2015, https://www.slideshare.net/ExtractConf/andrew-ng-chief-scientist-at-baidu (accessed September 9, 2021).

[32] Sanford J. Grossman and Oliver D. Hart, "The Costs and Benefits of Ownership: A Theory of Vertical and Lateral Integration," *Journal of Political Economy* 94 (4) (1986): 691–719; and Oliver D. Hart and John Moore, "Property Rights and the Nature of the Firm," *Journal of Political Economy* 98 (6) (1990): 1119–1158.

[33] Erik Brynjolfsson and Andrew Ng, "Big AI Can Centralize Decisionmaking and Power. And That's a Problem," MILA-UNESCO Working Paper (Montreal: MILA-UNESCO, 2021).

[34] "Simon Electronic Brain–Complete History of the Simon Computer," History Computer, January 4, 2021, https://history-computer.com/simon-electronic-brain-complete-history-of-the-simon-computer/.

[35] Hans *Moravec (1988), Mind Children,* Harvard University Press

[36]

Infinite Lives," *Econometrica* 54 (3) (1986): 607–622; and Kenneth L. Judd, "Redistributive Taxation in a Simple Perfect Foresight Model," *Journal of Public Economics* 28 (1) (1985): 59–83.

[47] Tambe et al., "Digital Capital and Superstar Firms."

[48] Katherine S. Newman, *Chutes and Ladders: Navigating the Low-Wage Labor Market* (Cambridge, Mass.: Harvard University Press, 2006).

[49] While the distinction between complements and substitutes is clear in economic theory, it can be trickier in practice. Part of the appeal of broad training and/or tax incentives, rather than specific technology mandates or prohibitions, is that they allow technologies, entrepreneurs, and, ultimately, the market to reward approaches that augment labor rather than replace it.

[50] See David H. Autor, David Dorn, and Gordon H. Hanson, "The China Shock: Learning from Labor-Market Adjustment to Large Changes in Trade," *Annual Review of Economics* 8 (2016): 205–240.